# Ground state tuning of the metal-insulator transition by compositional variations in BaIr$_{1-x}$Ru$_x$O$_3$ ($0 \leq x \leq 1$)

S. J. Yuan[1], K. Butrouna[1], J. Terzic[1], H. Zheng[1], S. Aswartham[1], L. E. DeLong[1], P. Schlottmann[2] and G. Cao[1]

[1] *Center for Advanced Materials and Department of Physics and Astronomy, University of Kentucky, Lexington, Kentucky 40506, USA*

[2] *Department of Physics, Florida State University, Tallahassee, FL 32306, USA*

## ABSTRACT

BaIrO$_3$ is a magnetic insulator driven by the spin-orbit interaction (SOI), whereas BaRuO$_3$ is a paramagnet and exhibits a crossover from a metallic to an insulating regime. Our investigation of structural, magnetic, transport and thermal properties reveals that substitution of Ru$^{4+}$ (4d$^4$) ions for Ir$^{5+}$ (5d$^5$) ions in BaIrO$_3$ reduces the magnitudes of the SOI and a monoclinic structural distortion, and rebalances the competition between the SOC and the lattice degrees freedom to generate a rich phase diagram for BaIr$_{1-x}$Ru$_x$O$_3$ ($0 \leq x \leq 1$). There are two major effects of Ru additions: (1) Light Ru doping ($0 < x \leq 0.15$) prompts simultaneous, precipitous drops in both the magnetic ordering temperature T$_N$ and the electrical resistivity, which exhibits a crossover behavior from a metallic to an insulating state near T$_N$. (2) Heavier Ru doping ($0.41 \leq x \leq 0.9$) induces a robust metallic state with a strong spin frustration generated by competing antiferromagnetic and ferromagnetic interactions.

## I. INTRODUCTION

A unique feature of the 5d-iridates is that a strong spin-orbit interaction (SOI) competes vigorously with Coulomb interactions, non-cubic crystalline electric fields, and Hund's rule coupling[1-5]. The relative strengths of these interactions stabilize new exotic ground states that provide a fertile ground for studying new physics. In particular, it is now recognized that strong SOI can drive novel, narrow-gap Mott insulating states in iridates. The SOI is a relativistic effect that is proportional to $Z^4$ ($Z$ is the atomic number), and is approximately 0.4 eV in the iridates (compared to ~ 20 meV in 3d materials), and splits the $t_{2g}$ bands into states with $J_{eff} = 1/2$ and $J_{eff} = 3/2$, the latter having lower energy. Since the $Ir^{4+}$ ($5d^5$) ions provide five 5d valence electrons, four of them fill the lower $J_{eff} = 3/2$ bands, and one electron partially occupies the $J_{eff} = 1/2$ band in which the Fermi level $E_F$ resides. The $J_{eff} = 1/2$ band is so narrow that even a reduced U (~ 0.50 eV, due to the extended nature of 5d-electron orbitals) is sufficient to open a gap ($\leq$ 0.62 eV) that induces a novel insulating state, which is contrary to expectations based upon the relatively large, unsplit $5d$ bandwidth[1-3,6].

Adopting a distorted hexagonal structure with both face-sharing and corner-sharing $IrO_6$ octahedra, $BaIrO_3$ is particularly unique, in that it exhibits a simultaneous onset of weak ferromagnetic (WFM, due to a canted antiferromagnetic structure) and charge density wave (CDW) orders with an unexpectedly high Néel temperature $T_N = 183$ K, and a temperature-driven transition from a bad-metal to an insulating ground state [7,8]. The ground state of $BaIrO_3$ is extremely sensitive to lattice contractions that can be tuned by light doping or the application of low hydrostatic pressures[4,9]. The extraordinary delicacy of the ground state in $BaIrO_3$ implies a critical balance between orbital, electronic, and lattice degrees of freedom [4]. The hexagonal structure of $BaIrO_3$ is similar to that of nine-layered, rhombohedral (9R) $BaRuO_3$, which exhibits a crossover from metallic to insulating behavior and enhanced paramagnetism with decreasing temperature [10,11]. However, a monoclinic distortion extant in $BaIrO_3$ at room temperature and 90 K generates twisting and buckling of the cluster trimmers (see Fig. 1) that give rise to two one-dimensional (1D) zigzag chains along the *c*-axis, and a two-dimensional (2D) layer of corner-sharing $IrO_6$ octahedra in the *ab* plane [7,9,12-14].

Although $BaIrO_3$ and $BaRuO_3$ have similar structures, they exhibit sharply contrasting physical properties, which underscores the critical role SOI and the lattice degrees of freedom can play in determining the ground state in iridates. In this work, we reduce the magnitudes of the SOI and the lattice distortion by substituting $Ru^{4+}$ ($4d^4$) for $Ir^{5+}$ ($5d^5$) in single-crystal $BaIr_{1-x}Ru_xO_3 (0 \leq x \leq 1)$. Ru substitution tunes the ground state by adding holes to the $t_{2g}$ bands, which lowers $E_F$ and moves the system away from the Mott instability toward a more robust metallic state at the expense of disorder scattering. However, in a metallic environment, disorder scattering is less relevant. The Ir/Ru1-O-IrRu3 bond angle increases with *x*, which strengthens the antiferromagnetic (AFM) superexchange between neighboring octahedra. Additionally, the SOI decreases while the Hund's rule coupling is enhanced on the Ru sites, hence further strengthening the competition between AFM and ferromagnetic (FM) interactions. As a result of this competition, the WFM transition at $T_N$ is suppressed with increasing Ru concentration, and spin frustration arises at intermediate temperatures due to increased competition between ferromagnetic and antiferromagnetic interactions.

## II. EXPERIMENTAL

Single crystals were grown using flux techniques described elsewhere [7]. The crystals are plate-like with hexagonal surfaces and a visible layered texture along the *c*-axis. Sample structures were determined using a Nonius Kappa CCD X-ray diffractometer at 90 K, and were refined by full-matrix, least squares using the SHELX-97 programs [15]. The standard deviations of all lattice parameters and interatomic distances are smaller than 0.1%. Chemical compositions of the single crystals were estimated using a combined unit of Hitachi/Oxford SwiftED 3000 for energy dispersive X-ray (EDX) spectroscopy. The magnetization M(T), the resistivity ρ(T), and the specific heat C(T), were measured between 1.7 K and 400 K using a

Quantum Design 7T SQUID Magnetometer and a Quantum Design 9T Physical Property Measurement System, respectively.

### III. RESULTS AND DISCUSSION

The *C2/m* (12) space group of BaIrO$_3$ features three face-sharing IrO$_6$ octahedra forming Ir$_3$O$_{12}$ trimers that are corner- and face-shared via IrO$_6$ octahedra (containing Ir1 and Ir3 sites) to form one-dimensional (1D) chains along the *c*-axis [9,11-14]. A monoclinic distortion generates twisting and buckling of the trimers (tilted ~ 12 ° relative to each other), which gives rise to two 1D zigzag chains along the *c*-axis, and a two-dimensional layer of corner-sharing IrO$_6$ octahedra in the *ab*-plane (see Fig. 1(a)). Substituting Ru$^{4+}$ for Ir$^{4+}$ preserves the crystal structure and results in a nearly uniform ~ 3% reduction in unit cell volume $V$ at $x = 0.63$, as shown in Table 1. This behavior is expected for Ru$^{4+}$ doping because the ionic radius of Ru$^{4+}$ (0.620 Å) is slightly smaller than that of Ir$^{4+}$ (0.625 Å). The *a*-axis is compressed by 0.5%, and the *c*-axis by 2.7%. In addition, the Ir/Ru1-O-Ir/Ru3 bond angle $\theta$ increases with Ru concentration and reaches 180 ° for x = 1 (i.e., BaRuO$_3$), indicating a less distorted lattice. It is already established that $\theta$ is critical to the electronic and magnetic structure of iridates [4]. 9R-BaRuO$_3$ exhibits a similar crystal structure with the $R\bar{3}m$ (166) space group, as shown in Fig. 1(b). Three RuO$_6$ octahedra share faces in a partial chain, facilitating direct Ru-Ru *d*-orbital interactions between the octahedra. Each of these triple units of octahedra shares corners with its neighbors along the hexagonal axis via nearly 180 ° Ru1-O-Ru3 bonds that favor superexchange coupling. A structural transition is to be expected somewhere in the composition range, 0.63 < x < 1.

Table 1: Lattice parameters, Ir/Ru1-O2-Ir/Ru3 bond length and angle at 90 K

| x | a (Å) | b (Å) | c (Å) | β (°) | Ir/Ru1-O (Å) | Ir/Ru3-O (Å) | Ir/Ru1-O-Ir/Ru3 (°) | Space Group |
|---|---|---|---|---|---|---|---|---|
| 0.0 | 9.9935 | 5.7352 | 15.2376 | 103.4111 | 2.0016 | 2.0153 | 161.560 | C2/m(12) |
| 0.1 | 9.9839 | 5.7377 | 15.1107 | 103.3402 | 1.9918 | 2.0132 | 163.678 | C2/m(12) |
| 0.63 | 9.9440 | 5.7429 | 14.8102 | 102.8574 | 1.9897 | 1.9731 | 174.296 | C2/m(12) |
| 1 | 5.7366 | 5.7366 | 21.5933 | NA | 1.9730 | 1.9730 | 180.0 | $R\bar{3}m$ (166) |

Ru doping induces pronounced changes in a wide range of physical properties of single-crystal $BaIr_{1-x}Ru_xO_3$. Representative data for χ(T) show the weak magnetic transition at $T_N$ is effectively depressed from 183 K for $x = 0$, to 145 K for $x = 0.04$, and vanishes for x ≥ 0.41, as shown in Fig. 2. The magnetic anisotropy also decreases with Ru additions, due to reduction of the SOI and increased Hund's rule exchange interaction. The magnetic data in Fig. 2 were fitted to a Curie-Weiss law $\chi = \chi_0 + C/(T - \theta_{CW})$ over the temperature range $180 \leq T \leq 300$ K for $0 \leq x \leq 0.9$ ($\chi_0$ is a temperature-independent constant, $\theta_{CW}$ the Curie-Weiss temperature, and $C$ the Curie constant). We then used $\chi_0$ to obtain $\Delta\chi = C/(T - \theta_{CW})$ and $\Delta\chi^{-1}$ vs $T$, as shown in the inset of Fig. 2(b). When $x = 1$ (i.e., $BaRuO_3$), the magnetic susceptibilities cannot be fitted to the Curie-Weiss expression, since the susceptibility increases with increasing temperature, which is an effect attributed to the unique quasi-one-dimensional structure [10]. The results of the fitting are shown in Figs. 3(b) and 3(c). Addition of Ru changes the sign of $\theta_{CW}$ from positive for the parent compound, $BaIrO_3$, to negative through most of the composition range, indicating that the average exchange coupling shifts from FM to AFM. The increase in the absolute value of $\theta_{CW}$ is an indication of the increase of the AFM exchange interaction strength with increasing $x$. As shown in Fig. 3(a), the Ir/Ru1-O-Ir/Ru3 bond angle $\theta$ increases with Ru concentration and reaches 180 ° for $x = 1$. The increase of the bond angle $\theta$ directly enhances the AFM interaction between Ir/Ru1 and Ir/Ru3. Since $\theta_{CW}$ measures the strength of the magnetic interaction, a large absolute value of $\theta_{CW}$ in a system without magnetic ordering above 1.7 K implies a strong spin frustration. It is conceivable that both the disappearance of magnetic order at $x = 0.41$ and the gradual appearance of spin frustration are consequences of competing AFM and FM interactions resulting from atomic disorder on the Ru and Ir sites. This changes the local energies with $x$, such as a decrease of the SOI, modifies the noncubic CEF, enhances the Hund's rule exchange interaction, and intensifies the competition between FM and AFM correlations. The mismatch of the $t_{2g}$ energy levels also reduces the tight-binding hopping of the 4d/5d electrons, leading to more localized states and hence an enhanced $\mu_{eff}$.

Ru doping also strongly impacts the transport properties, as indicated in Fig. 4. The temperature dependence of the resistivity in both the *ab*-plane and *c*-axis, $\rho_{ab}(T)$ and $\rho_c(T)$, respectively, exhibit a sharp kink at $T_N = 183$ K for $x = 0$, consistent with previous results [7,12]. This transition has been associated with the formation of a CDW gap ($E_g \sim 0.15$ and $0.13$ eV for current within the *ab*-plane and along the *c*-axis, respectively) that is a precursor to BaIrO$_3$ developing an insulating state below $T_N$. BaIrO$_3$ becomes an insulator only at low T; and the gap can be readily reduced by merely a few percent of Ru doping. The Ru impurity states gradually fill the gap, which rapidly becomes a pseudogap. This leads to an insulator-metal transition as a function of *x*, as seen in Figs. 4(a) - 4(f). The electrical resistivity $\rho_c(T)$ is reduced by more than two orders of magnitude (from 135 m$\Omega$-cm at $x = 0$ to 0.75 m$\Omega$-cm) for T = 1.8 K and $x = 0.04$. Indeed, dilute Ru substitutions for Ir immediately result in a reduced $\rho(T)$ and a metallic temperature dependence at high temperatures. The metallic behavior of $\rho(T)$ clearly becomes stronger with increasing *x*. There is a kink with a minimum resistivity value at T = 135 K for $x = 0.04$, which corresponds to the transition to weak magnetic order at $T_N$, as shown in Fig. 2(a). The resistivity displays metallic behavior with $d\rho/dT > 0$ above T = 135 K, and $\rho_c(T)$ and $\rho_a(T)$ are radically reduced by three orders of magnitude below 135 K. On the other hand, the resistivity exhibits a noticeable upturn as the temperature decreases, which indicates that a low-temperature metallic state is not yet fully realized. The energy gap values obtained from fitting to an activation law indicate that the insulating gap is not fully closed (see inset of Fig. 4(a)) for $x = 0.15$, for which the resistivity values are generally larger than for $x = 0.04$, but the metal-insulator crossover behavior is not apparent, possibly due to strong disorder scattering. It is noted that $\rho_a(T)$ for $x = 0.15$ roughly follows a variable range hopping (VRH) behavior, $\rho \sim \exp(1/T^{1/2})$, below 50 K (see the inset in Fig. 4(c)), which implies that Anderson localization comes into play at $x = 0.15$. When $x \geq 0.41$, Ru substitution adds more holes to the bands, which lowers $E_F$ and moves the system away from the Mott instability toward a robust metallic state. Disorder scattering is then less relevant. For $x = 1$ (i.e., BaRuO$_3$), a crossover from metallic to insulating behavior reappears at low temperature, resulting from pseudogap formation and 1D-CDW fluctuations [11].

The temperature dependence of the specific heat $C$ for various $x$ is given in Fig. 5(a). Fitting the data to $C(T) = \gamma T + \beta T^3$ for $7 < T < 17$ K yields the Sommerfeld coefficient $\gamma$ for the electronic contribution to $C(T)$ (see Fig. 5(b)), which serves as a measure of the electronic density of states at the Fermi level, $N(E_F)$, and the effective mass of the carriers. There is a substantial increase of $\gamma$ with dilute Ru concentration; in particular, $\gamma$ reaches 11.75 mJ/mol K$^2$ for $x = 0.04$, and 15.09 mJ/mol K$^2$ for $x = 0.15$, compared to $\gamma = 2.34$ mJ/mol K$^2$ for the parent compound ($x = 0.0$). This value for $x = 0.0$ is rather large and is possibly due to localized states arising from defects. The increase of $\gamma$ initially reflects the increase of $N(E_F)$ induced by hole doping by Ru ions. Nevertheless, $N(E_F)$ and $\gamma$ eventually decrease with $x$, as shown in Fig. 5(b). In the case of BaRuO$_3$, the smaller values reflect pseudogap formation due to the CDW instability [11].

The central findings of this study are summarized in Fig. 6, which shows a phase diagram for BaIr$_{1-x}$Ru$_x$O$_3$. The initial Ru doping reduces the SOI, and alters the relative strength of the SOI and the tetragonal CEF that dictate the magnetic state, which, in turn, affects the band gap near E$_F$. Ru doping also enhances the Hund's rule coupling that competes with the SOI and prevents the formation of the J$_{eff}$ = 1/2 state [5,16]. In addition, Ru doping relaxes the crystal structure and the Ir/Ru1-O-Ir/Ru3 bond angle $\theta$ increases with Ru doping, which enhances the AFM coupling between neighboring octahedra. These changes account for the precipitous decrease in T$_N$, which vanishes around $x = 0.15$. Ru additions introduce holes to the conduction band and disorder, and give rise to a increased N(E$_F$). Hence, a small amount of Ru doping leads to a largely reduced magnitude of $\rho(T)$. An energy level mismatch between the Ru and Ir sites makes the hopping of the carriers between octahedra containing differing $d$-elements more difficult, and also changes the orientation angles of the octahedra. The random Ru/Ir occupation gives rise to Anderson localization and an insulating state for $0.15 \leq x < 0.41$. As $x$ increases further ($0.41 \leq x < 1$), the addition of holes induces a more robust metallic state, and disorder scattering is of lesser importance. The reduction of the Ir/Ru1-O-IrRu3 bond angle favors superexchange and AFM coupling between neighboring octahedra. Moreover, the SOI decreases while the Hund's rule coupling is enhanced (on the Ru sites), which further strengthens the competition between AFM and FM couplings, which gives rise to spin frustration at intermediate temperatures.

**ACKNOWLEDGMENTS**

This work was supported by the National Science Foundation via Grant No. DMR-1265162 (GC), Department of Energy (BES) through grants No. DE-FG02-98ER45707 (PS) and DE-FG02-97ER45653 (L.E.D.).

**REFERRENCES**

[1] B. J. Kim *et al.*, Physical Review Letters **101**, 076402 (2008).

[2] B. J. Kim, H. Ohsumi, T. Komesu, S. Sakai, T. Morita, H. Takagi, and T. Arima, Science **323**, 1329 (2009).

[3] G. Cao and L. E. DeLong, in *Frontiers of 4d- and 5d-Transition Metal Oxides* (World Scientific, Singapore, 2013).

[4] O. B. Korneta, S. Chikara, S. Parkin, L. E. DeLong, P. Schlottmann, and G. Cao, Physical Review B **81**, 045101 (2010).

[5] T. F. Qi, O. B. Korneta, L. Li, K. Butrouna, V. S. Cao, X. Wan, P. Schlottmann, R. K. Kaul, and G. Cao, Physical Review B **86**, 125105 (2012).

[6] J. Dai, E. Calleja, G. Cao, and K. McElroy, Physical Review B **90**, 041102 (2014).

[7] G. Cao, J. E. Crow, R. P. Guertin, P. F. Henning, C. C. Homes, M. Strongin, D. N. Basov, and E. Lochner, Solid State Communications **113**, 657 (2000).

[8] M. L. Brooks, S. J. Blundell, T. Lancaster, W. Hayes, F. L. Pratt, P. P. C. Frampton, and P. D. Battle, Physical Review B **71**, 220411 (2005).

[9] G. Cao, X. N. Lin, S. Chikara, V. Durairaj, and E. Elhami, Physical Review B **69**, 174418 (2004).

[10] M. Shepard, S. McCall, G. Cao, and J. E. Crow, Journal of Applied Physics **81**, 4978 (1997).

[11] Y. S. Lee, J. S. Lee, K. W. Kim, T. W. Noh, J. Yu, Y. Bang, M. K. Lee, and C. B. Eom, Physical Review B **64**, 165109 (2001).

[12] A. V. Powell and P. D. Battle, Journal of Alloys and Compounds **191**, 313 (1993).

[13] M. H. Whangbo and H. J. Koo, Solid State Communications **118**, 491 (2001).

[14] R. Lindsay, W. Strange, B. L. Chamberland, and R. O. Moyer Jr, Solid State Communications **86**, 759 (1993).

[15] G. Sheldrick, Acta Crystallographica Section A **64**, 112 (2008).

[16] H. Watanabe, T. Shirakawa, and S. Yunoki, Physical Review Letters **105**, 216410 (2010).

**Figure Captions**

FIG. 1. Crystal structure of (a) BaIrO$_3$ and (b) BaRuO$_3$. Note the corner-sharing Ir$_3$O$_{12}$ and Ru$_3$O$_{12}$ trimers that are connected through the vertices of the top and bottom octahedra of the trimers, and the schematic of the M$_1$–O$_2$–M$_3$ bond angle (M = Ir or Ru).

FIG. 2. The magnetic susceptibilities χ(T) at $\mu_0 H = 0.1\ T$ for BaIr$_{1-x}$Ru$_x$O$_3$, where (a) $0 \leq x \leq 0.15$ and (b) $0.42 \leq x \leq 1$. The inset in (b) shows χ(T) for $x = 0.04$ and 0.1.

FIG. 3. The Ru concentration $x$ dependence of (a) the Ir/Ru$_1$–O$_2$–Ir/Ru$_3$ bond angle, (b) $T_N$ and $\theta_{CW}$, and (c) the magnetic effective moment $\mu_{eff}$.

FIG. 4. The temperature dependence of the resistivity ρ(T) for BaIr$_{1-x}$Ru$_x$O$_3$. The inset in (a) illustrates activated behavior and the associated energy gaps. The inset in (c) illustrates variable range hoping (VRH) in a plot of $\ln\rho_a$ vs $T^{-1/2}$. The vertical arrows indicate the kink that corresponds to the weak magnetic transition at $T = T_N$.

FIG. 5. (a) The specific heat $C(T)/T$ vs $T^2$, and (b) the Sommerfeld coefficient $\gamma$ vs $x$, for BaIr$_{1-x}$Ru$_x$O$_3$.

FIG. 6. Phase diagram for BaIr$_{1-x}$Ru$_x$O$_3$ based upon data presented herein. WFM-I stands for weak ferromagnetic insulator, PM-I stands for paramagnetic insulator, and PM-M stands for paramagnetic metal.

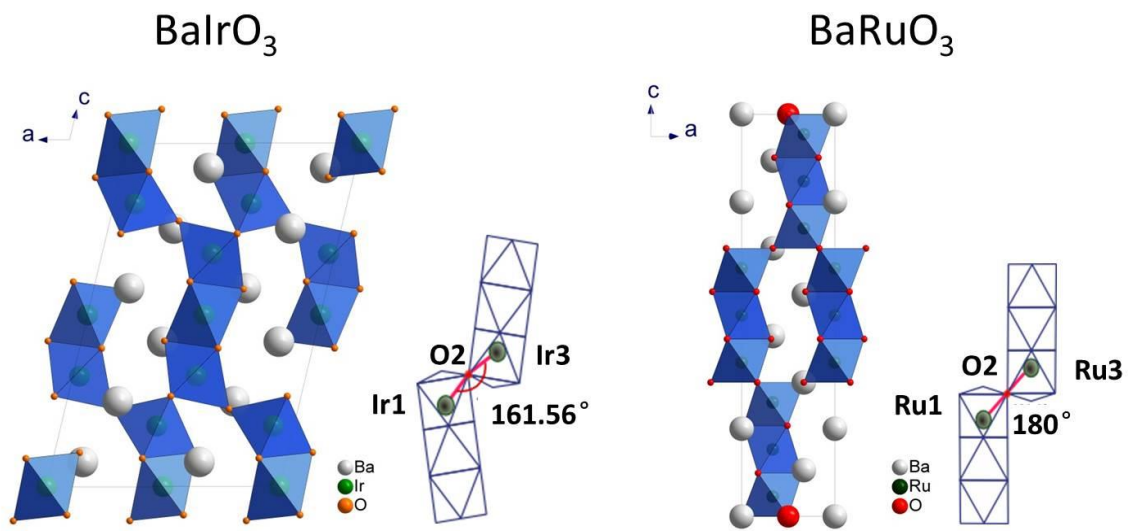

Fig.1

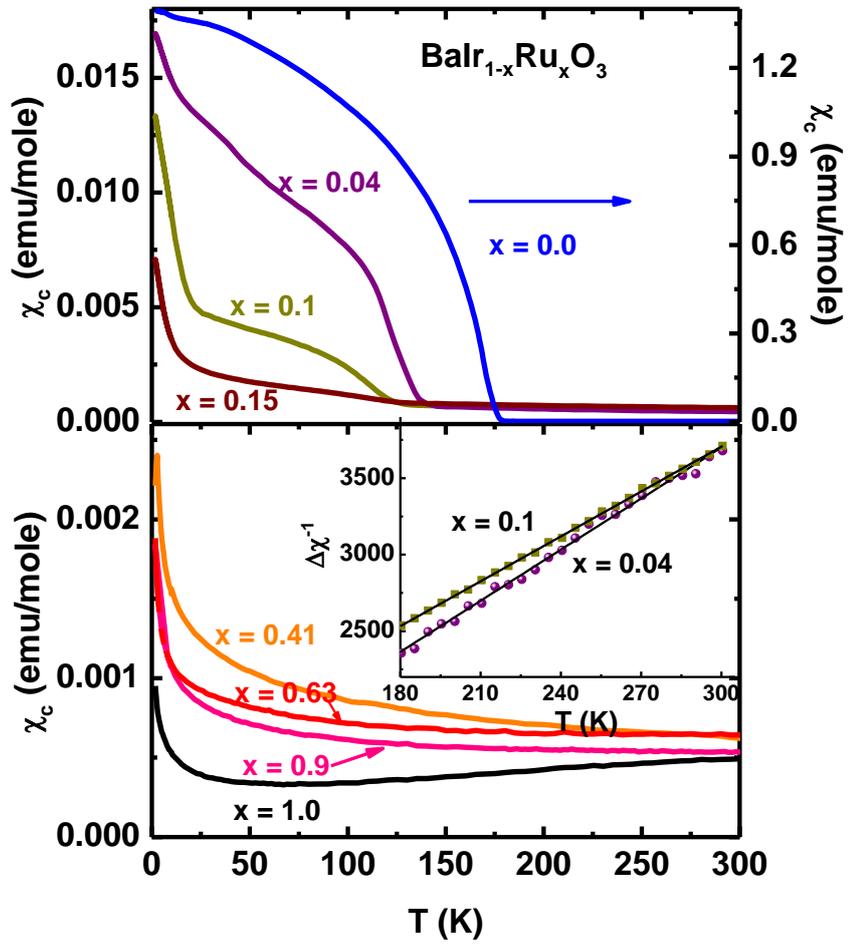

Fig.2

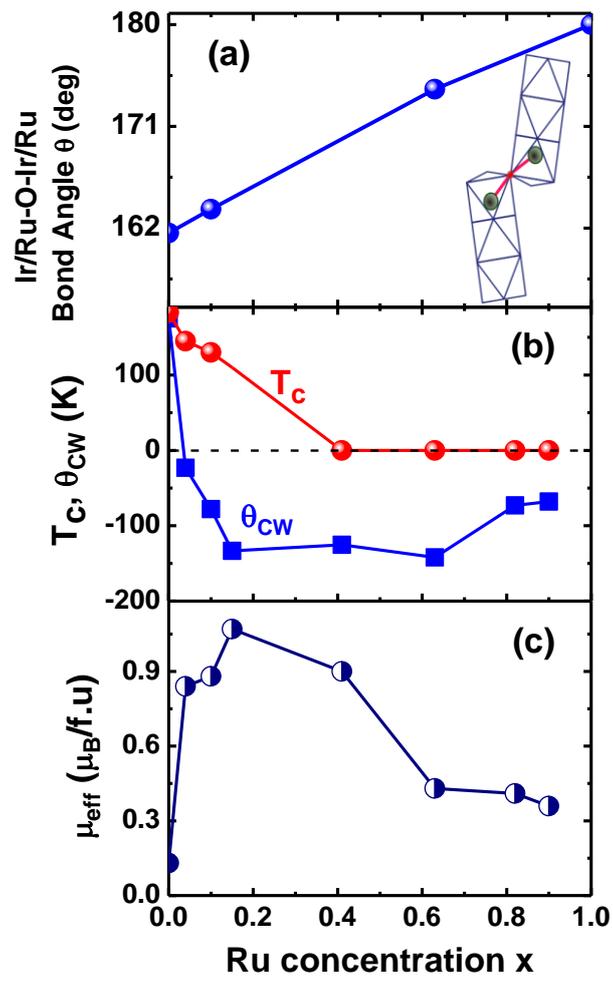

Fig. 3

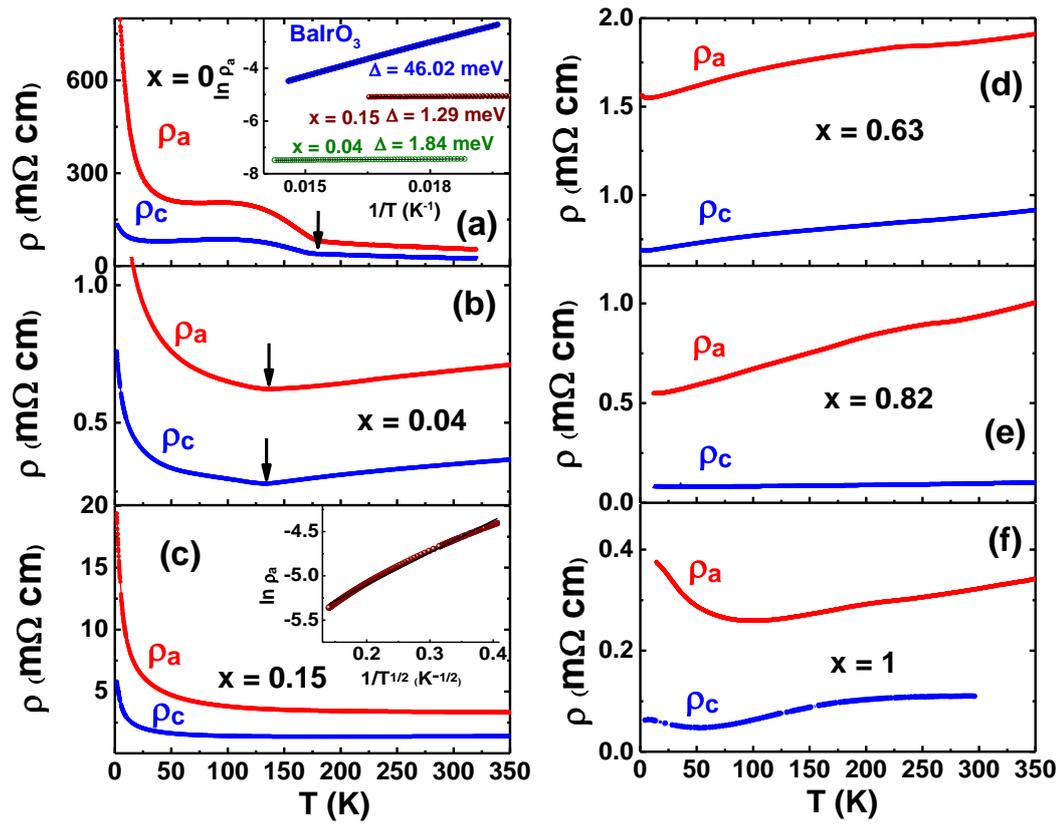

Fig.4

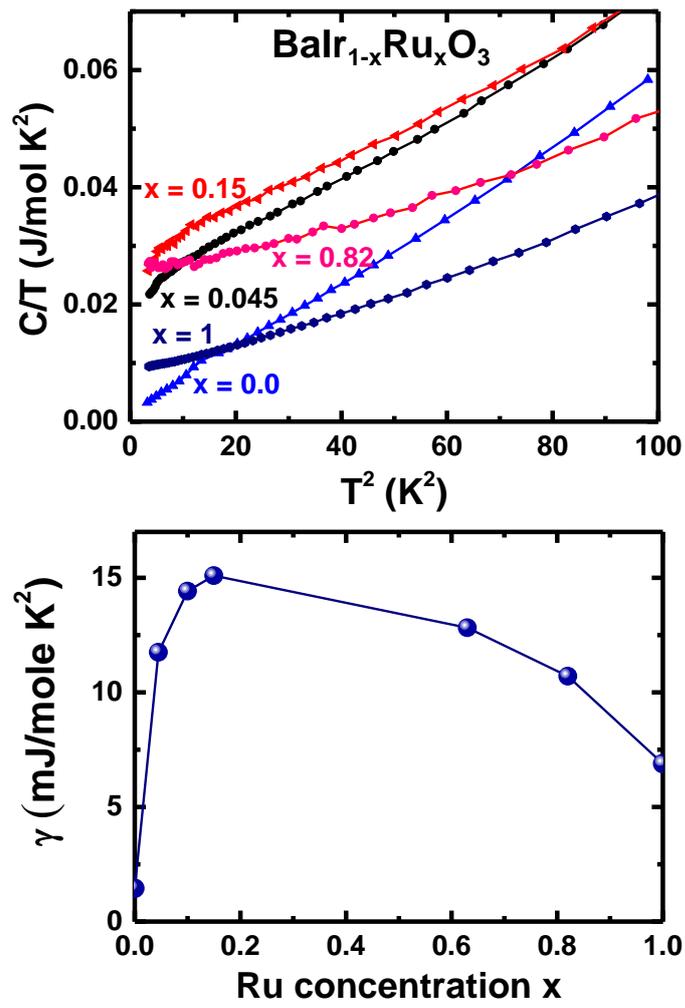

Fig.5

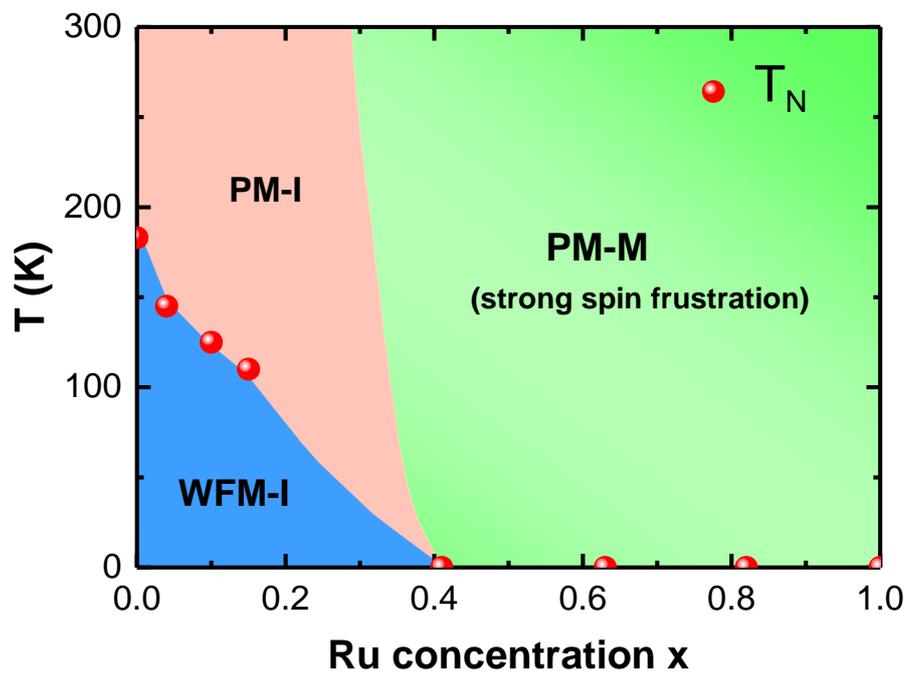

Fig.6